# Effects of Sensemaking Translucence on Distributed Collaborative Analysis

**Nitesh Goyal**
Dept of Information Science
Cornell University
ngoyal@cs.cornell.edu

**Susan R. Fussell**
Dept of Information Science
Cornell University
sfussell@cornell.edu

**ABSTRACT**

Collaborative sensemaking requires that analysts share their information and insights with each other, but this process of sharing runs the risks of prematurely focusing the investigation on specific suspects. To address this tension, we propose and test an interface for collaborative crime analysis that aims to make analysts more aware of their sensemaking processes. We compare our sensemaking translucence interface to a standard interface without special sensemaking features in a controlled laboratory study. We found that the sensemaking translucence interface significantly improved clue finding and crime solving performance, but that analysts rated the interface lower on subjective measures than the standard interface. We conclude that designing for distributed sensemaking requires balancing task performance vs. user experience and real-time information sharing vs. data accuracy.

**ACM Classification Keywords**
H.5.3. Information interfaces and presentation (e.g., HCI): Groups and Organization Interfaces: Computer-supported cooperative work

## INTRODUCTION

In March 2008, Demetrius Smith was charged for the murder of Robert Long. Even though Long was working with police as an informant and potential witness against his boss named Morales, police ignored Morales as a potential suspect. Despite Morales having motive, and opportunity, they decided to pursue incriminating Smith. It was not until April 2011 that the case was reopened because evidence pointed to racial information playing an important role in fake testimonies and police investigation. After serving a five year prison sentence, Smith was exonerated and released. The biased perception held by investigators hindered the process of sensemaking in two ways. First, the investigators should not only have collected evidence that confirmed their (wrong) hypothesis that Smith committed the crime, but also collected evidence that disconfirmed their hypothesis. Second, self-awareness of personal biases is hard. It is even harder in the process of complex sensemaking like crime analysis. In retrospect, awareness of biases might have afforded investigators the cognizance that their attention was prematurely focused on a single suspect instead of appropriately distributed across other suspects, including Morales. Thus, an absence of due process and transparency into one's own mental process enabled biased sensemaking.

Unfortunately the Smith case is not the only criminal case in which biases hinder sensemaking. Police Chief periodical reports that on average, 16 murders occur every day that might never be solved and their perpetrators never arrested because of reasons like confirmation biases and groupthink [43]. These issues may be exacerbated in cases where crime investigators in multiple agencies need to work together, due to reduced information sharing and awareness across geographically distributed teams and investigating partners [13, 37]. While the timely exchange of information is essential to successfully solving crimes, at the same time, information received from one analyst can unduly influence another's reasoning, resulting in cognitive tunneling as in the Morales case.

In the current work, we focus on the notion of *sensemaking translucence*, or the process of making analysts more aware of their sensemaking processes. Sensemaking involves foraging for information pieces that could connect with each other, resulting in multiple initial hypotheses. These hypothesis are then closely synthesized to find evidence that confirms or disconfirms them, until an ultimate hypothesis remains [42]. Successful crime investigators pursue multiple suspects until they have sufficient information to rule out all but one of them, the correct one [30]. While, the sensemaking process can go wrong when information is not shared in a timely fashion, it can also go wrong when an analyst prematurely decides on a suspect without ruling out the others as in the case of Demetrius Smith.

To balance the need for information exchange with the goal of reducing cognitive biases, we propose a sensemaking translucence interface that consists of two integrated parts: a *hypothesis window* that is intended to motivate explicit interchange of ideas about suspects' means, motives and alibis and a *suspect visualization* that provides automatic feedback on which suspects have been discussed based on the hypothesis window, a group chat window, and a digital sticky note feature. As we discuss in detail later, the design of the suspect visualization is intended to provide awareness not only of those suspects that have been discussed but also of the idea that there might be other suspects out there that have yet to be discussed.

We examine the effects of our sensemaking translucence interface in a laboratory study in which pairs of remote participants role-played detectives collaborating to solve a serial killer crime [2]. Half of the pairs used the sensemaking translucence interface and the other half used a standard interface [22] used in previous studies [19, 21]. As we will show, pairs using the sensemaking translucence interface were significantly better than those using the standard interface at uncovering pertinent clues and identifying the serial killer. However, participants viewed the sensemaking translucence interface as less valuable than the standard interface in terms of helping them focus their attention, develop hypotheses, or collaborate with their partners. The findings suggest that designers of collaborative analysis interfaces may want to incorporate real-time feedback to users about how the features of the interface are beneficial to their sensemaking processes.

In the remainder of the paper, we first discuss related work on the process of crime solving and the tools that have been designed to support this process, and outline our study hypotheses. We then present the design of the social translucence interface, the study methodology and measures, and the results. We conclude with a discussion of the contributions of this work to the development of tools to support collaborative analysis.

**RELATED WORK**

**Sensemaking in Collaborative Analysis**
Successful crime solving requires investigators to parse data, identify pertinent information, find potential suspects, and eventually identify the criminal [18]. This process of sensemaking is described by Pirolli and Card [42] as an iterative process of foraging and sensemaking. Analysts iteratively forage clues and generate mental models that represent the best explanation of what they have found, often generating multiple competing hypotheses before they finally choose the best explanation [30].

Collaboration in the sensemaking processes can be advantageous since partners can leverage each other's cognition and insights to solve hard problems [19, 27, 50]. Multiple analysts may have different access to documents, and with more readers there is a greater ability to sift through large amounts of data and identify patterns.

At the same time, collaborative sensemaking is challenging because analysts are often reluctant to exchange information and insights for fear they might be wrong [30]. This is a legitimate fear: the exchange of incorrect information can lead to poorer outcomes due to what Kang and Kiesler have termed *teammate inaccuracy blindness* [34]. That is, analysts treat all information from a partner as valid and useful, regardless of its actual quality. As a result, if one member of a team of analysts prematurely focuses in on an incorrect suspect, the other(s) is likely to follow.

Thus, the challenge for designers of collaborative analysis systems is to facilitate the positive benefits of information sharing and collaborative reasoning while preventing the negative consequences of biased analysis and incorrect solutions. We next consider existing tools in this design space and discuss how combinations of interface features may allow us to manage these two competing goals.

**Tools to Support Collaborative Sensemaking**
Leveraging partners' insights requires sharing of insights and subsequent awareness of these insights. Shared workspaces have been shown to improve shared understanding and awareness [17, 25] by promoting exchange of information and data with others [27], improving common ground [50] and increasing awareness of the status of the analysis task and others' activities in the task [11, 41]. Shared workspaces have primarily been researched from an explicit sharing perspective where analysts consciously choose to share their mental models and insights [7, 11, 39]. However, when asked to share explicitly, analysts often choose not to share until they are confident in their insights rather than sharing these insights at the right point in the collaborative sensemaking cycle.

Other approaches to supporting collaborative analysis include reminding analysts to view their partners' analysis, as in AnalyticStream [39] or recommending relevant pieces of information from their partner [5]. However, biases owing to personal beliefs may inhibit taking advantage of such features [e.g., 9, 33, 36] or lead to groupthink and/or cognitive tunneling [50], as in the case of Demetrius Smith.

More recently, implicit sharing of insights in two-person teams, where systems automatically share partners' generated information, has been shown to improve user experience significantly but provide limited gains in task performance [19]. Even with implicit sharing, analysts often resort to explicit channels of communication to continuously monitor the status of collaborative sensemaking cycle.

Other tools aim to help collaborators achieve common ground using collaborative visualizations [1, 8, 28, 31, 32, 46, 48]. Such visualizations can help users aggregate and abstract activities [28]. For example, users may send notifications of each activity [4], or be provided with a view into the dataset using a shared network diagram [1], a timeline [16], or a user activities list [29]. As with the other tools described above, analysts using visualization tools often need additional explicit communication through chats or comment threads [29] or annotations [35] to build up a shared mental model of the case.

### The Sensemaking Translucence Interface

To facilitate the exchange of insights while simultaneously discouraging cognitive tunneling, we developed what we call a *sensemaking translucence interface*. This interface consists of two main features: a hypothesis window and a suspect visualization.

The *hypothesis window* is similar to Alternative Competing hypothesis (ACH) [10], in which users explicitly share their hypotheses and evidence to maintain awareness of one another's insights and to develop a joint mental model of the case. The hypothesis window is also designed to help reduce confirmation bias by including fields for reporting evidence that disconfirms each hypothesis [2].

The *suspect visualization* depicts the joint attention paid to each suspect thus far in the analysis and encourages collaborators to distribute their attention across multiple suspects instead of focusing prematurely on a single suspect who might not be the actual culprit. The suspect visualization changes automatically as analysts mention suspects in their hypotheses, notes or chat conversations.

### Study Hypotheses

The hypothesis window and suspect visualization are designed to be used in tandem, such that each new sharing of a hypothesis is associated with steps to assess the quality of that hypothesis (via fields in the hypothesis window) and steps to promote consideration of other possible hypotheses (via the suspect visualization). We thus tested a sensemaking translucence interface that contained these integrated features against an earlier version of the same tool that did not [19]. For the reasons outlined above, we predicted that the sensemaking translucence interface would improve pairs' crime-solving performance:

H1. Participants using a sensemaking translucence interface will perform better on a collaborative analysis task than participants using a standard interface.

We also reasoned that by enabling analysts with a better understanding of their partners' thoughts and activities, the sensemaking translucence interface would help analysts make appropriate decisions about their own activity [15] and that analysts would perceive the sensemaking translucence interface to be of more value for their work than the standard interface.

H2a. Participants using a sensemaking translucence interface will rate the usefulness of the tool higher than participants using a standard interface.

H2b. Participants using a sensemaking translucence interface will report higher level of activity than participants using a standard interface.

We also believe that a sensemaking translucence interface has the potential to improve the experience of working together. Awareness of other analysts' activities has been

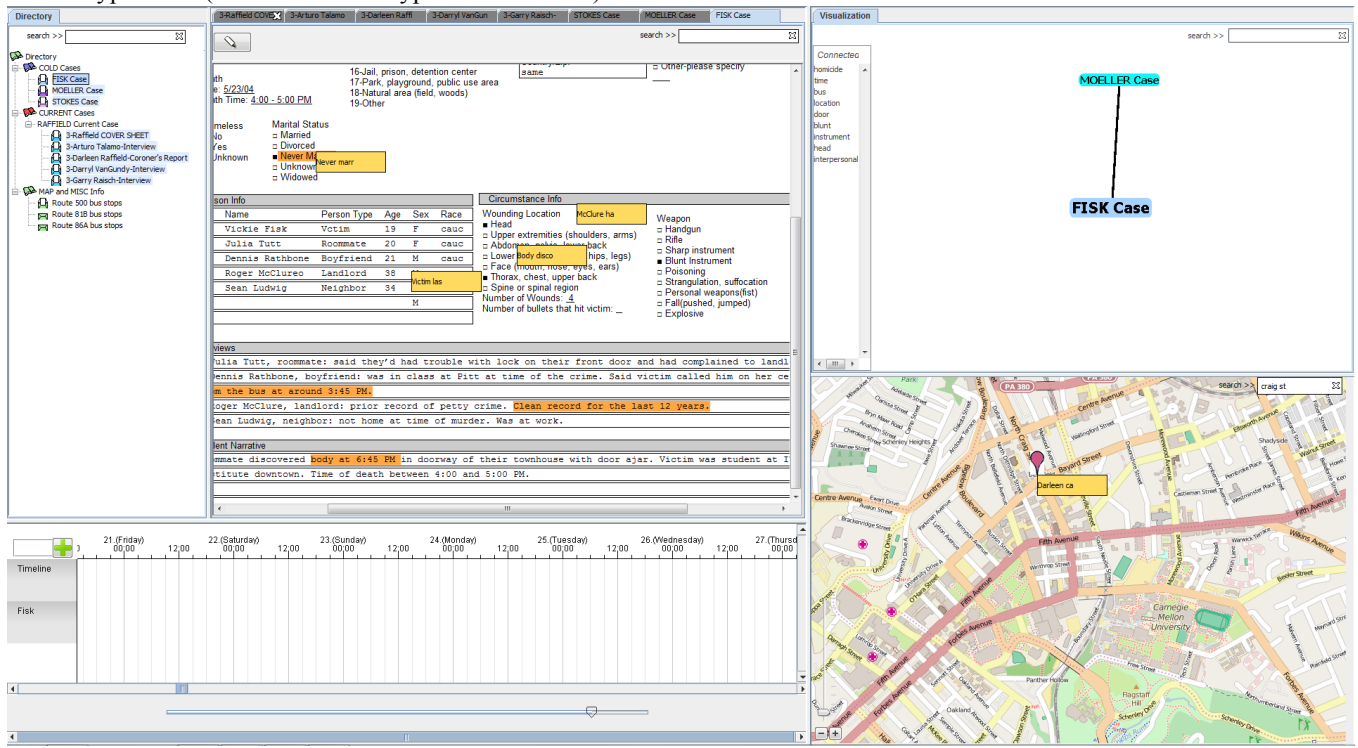

**Figure 1. The Document Space showing (clockwise, from top-left) the directory of crime case documents, a tabbed reader pane for reading case documents, a visual graph of connections based on common entities in the dataset, a map to identify locations of crimes and events, and a timeline to track events.**

shown to help novice analysts get up to speed [5]. In medical settings, implicitly shared awareness information can help establish common ground between clinical staff [3, 40] and lead to more positive perceptions of the process [10]. Since making sensemaking more transparent reduces uncertainty about the status of the task and reduces the need for verbal updates of status via the chat interface, we predicted:

H3. Participants using a sensemaking translucence interface will rate their collaborative experience higher than participants using a standard interface.

However, sensemaking translucence may also come with costs. Analysts may feel compelled to share preliminary thoughts, and read their partners' emergent hypotheses. This may increase the cognitive demand of the crime-solving task. On the other hand, by reducing the need for explicit verbal sharing of information, our interface may reduce the time and effort required for the task [14, 47]. There is also a potential for the suspect visualization to be distracting. Since the direction of impact is unclear, we pose a research question:

RQ1. How will the sensemaking translucence interface affect participants' cognitive workload?

## METHOD

We report data from an experiment in which pairs collaborated to identify a pattern in a crime dataset. They worked on solving these crimes using a simulated geographically distributed environment. Pairs were randomly assigned to one of two interface conditions: standard interface or sensemaking translucence interface. The *Standard Interface* included a document space and an analysis space where users could share information using stickies and chat. The *Sensemaking Translucence Interface* allowed users to share information like the standard interface and further enabled partners to track the progress of their analysis by explicit hypothesis tracking and suspect tracking visualization. We measured task performance, perceptions of the interface, quality of the collaborative experience and cognitive load.

### Research Prototype Tool

SAVANT, the prototypical tool, used for this experiment is based on previous work by Goyal et al. [19, 21, 22]. The current SAVANT has two main components: a *Document Space* and an *Analysis Space*.

The *Document Space* (Figure 1) was identical for both the standard interface and the sensemaking translucence interface. Here, investigators could view their case documents, and highlight/annotate text in these documents. They could also view and manipulate a network diagram that showed connections between cases as calculated by TF/IDF on named entities, access and annotate Google Maps to mark crime-locations, and annotate a timeline to identify temporal patterns. The Document Space appeared on one of the analysts' two monitors.

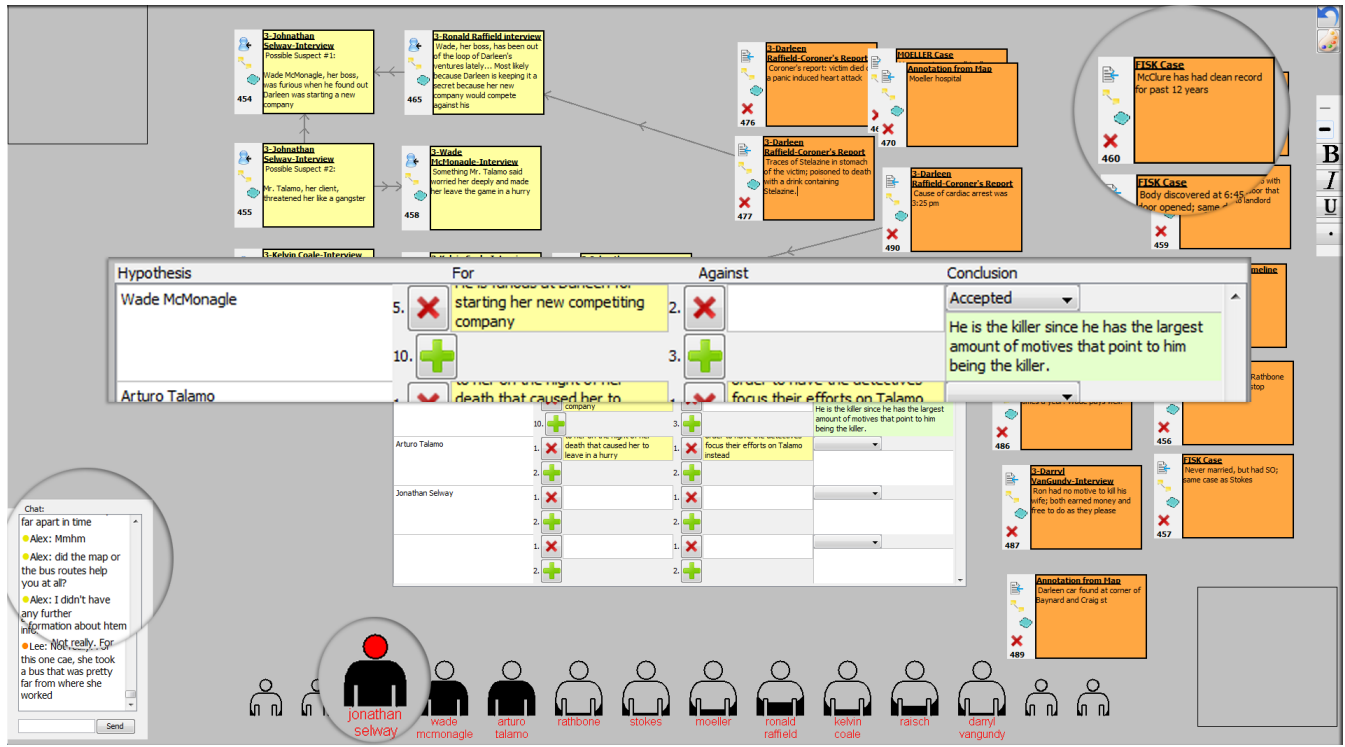

**Figure 2. The Analysis Space showing (clockwise, from top-left) the chat for explicit sharing, connected stickies for Implicit Sharing, hypothesis window in the middle with columns to add new Hypotheses, confirming evidence, and disconfirming evidence for explicit Hypothesis Tracking, and Suspect Visualization at the bottom with 4 Avatars: Dennis Rathbone. Marilyn Stokes, Steve Gramming, and Lousie for Suspect Tracking. Note: Chat, Visualization, Sticky, and Hypothesis Window have been magnified to improve readability.**

A second monitor was used to present the *Analysis Space* (Figure 2). Two features of the Analysis Space were common to participants in both the Standard Interface and the sensemaking translucence interface: digital stickies and a chat box. Annotations created in the Document Space appeared automatically as digital stickies in the Analysis Space, where they could be moved, edited, connected using arrows, or piled atop one another to show relevance. This iterative reorganization of stickies supports analysts' processes of foraging and sensemaking [6, 11, 27, 41]. The Analysis Space also included a standard chat box (Figure 1b, lower left).

The Analysis Space for participants in the sensemaking translucence interface included two additional features: a *Hypothesis Window*, and a *Suspect Visualization*. These two features are connected to each other, and enable sensemaking translucence in two different ways.

The *Hypothesis Window* (Figure 1b, center) allows users to enter their emergent hypotheses manually, reflecting on their current cognitive state of sensemaking. This space also reminds users to add evidence that confirms and disconfirms these hypotheses, such that users can explicitly mark the status of each hypothesis as accepted, rejected, or needing more information. Entries (hypothesis, confirming/disconfirming evidence, status, and status related comments) were color coded to reflect each team member's contribution.

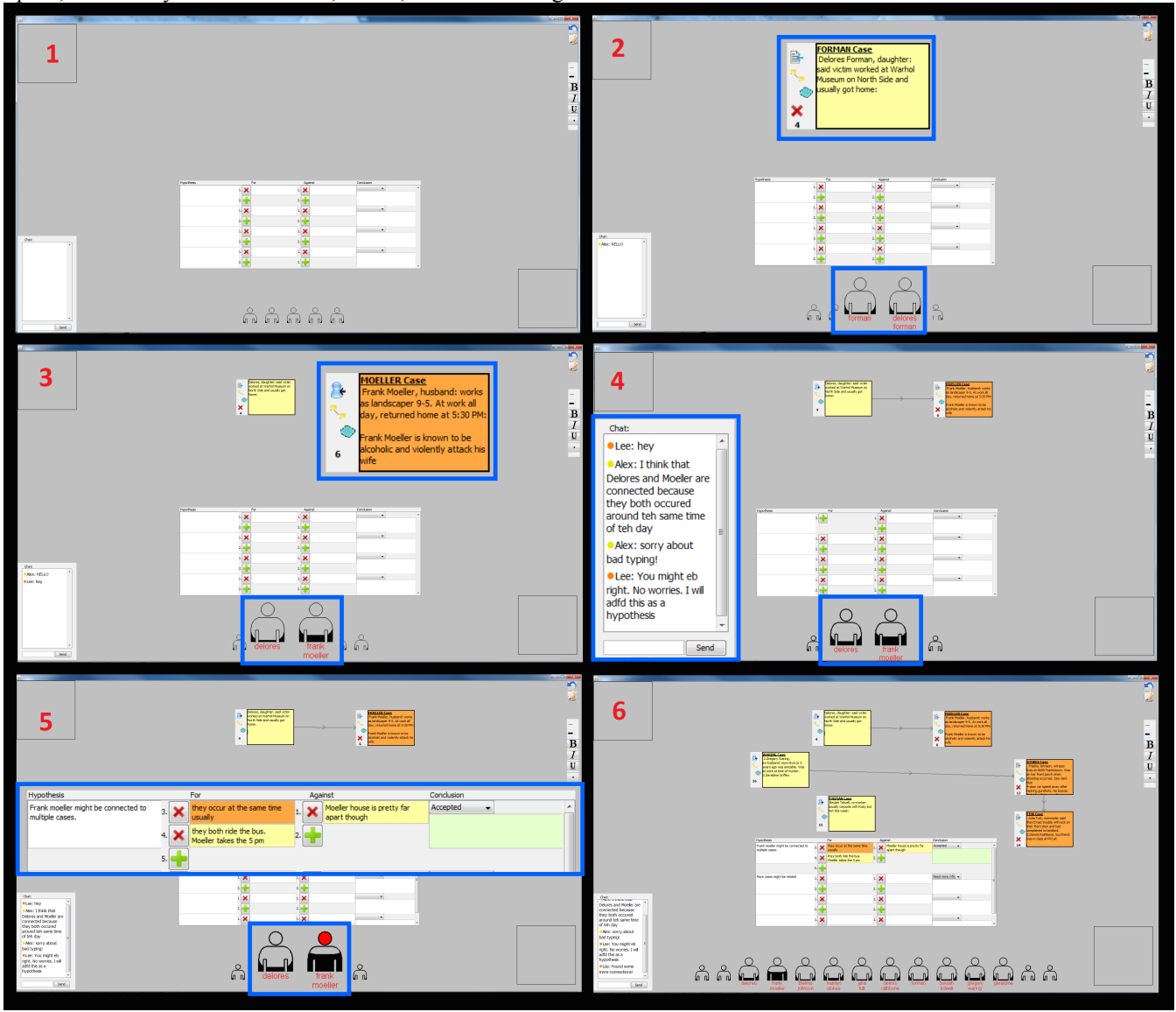

**Figure 3. Sample sensemaking trajectory 1. Sensemaking-Translucence reminds users to consider suspects by showing empty Avatars at the start 2. Avatars are automatically populated by names detected from implicitly shared stickies. 3, 4 & 5. Avatars show distribution of name-reference by getting darker for names mentioned in stickies, chat and hypothesis window. The last mentioned suspect in hypothesis window is marked red 6. With use, visualization depicts distribution of attention at suspect level, based on explicit mentions. Note: Chat, Visualization, Sticky, and Hypothesis Window have been magnified to improve readability in 2, 3, 4 and 5. 1 and 6 represent non-magnified versions of the Sensemaking Translucence interface.**

The *Suspect Visualization* (Figure 2, bottom center) was generated by the SAVANT system in real time using Natural Language Processing of named entities. The system automatically identified named entities (names of persons only) in stickies, the chat conversation, and the hypothesis window. Each newly identified name was assigned an avatar. The visualization begins with four unnamed avatars, suggesting that users should pursue names of potential suspects while the potential suspect-space is empty. As users share more suspect names in the Analysis Space, newly created named Avatars flanked by unnamed Avatars further remind users that there may be more suspects to discover. Furthermore, each time a name is mentioned in the Analysis Space, the associated Avatar darkens. This reflects the lack of non-equitable distribution of information sharing in the Analysis Space and supports suspect tracking. Figure 3 shows one possible sensemaking trajectory where the two sensemaking translucence features facilitated the exchange of insights while simultaneously discouraging cognitive tunneling.

In summary, there were two different versions of the SAVANT interface. In the standard interface condition, participants had the Document Space and an Analysis Space with stickies and chat box. In the sensemaking translucence interface condition, participants had the Document Space and an Analysis Space that included the Hypothesis Window and Suspect Visualization in addition to stickies and a chat box.

**Participants**
Fifty participants participated in the experiment described as a "Solve Crimes Together Study" as 25 pairs. Of the 25 pairs, data for five pairs was discarded due to technical failures in Internet connectivity (4 pairs) and inconsistent instructions (1 pair). Finally, forty participants participated in the experiment (16 male, 24 female; 77.5% U.S. born; age range 18-28, median age approximately 21; 82.5% spoke English as first language). All students were undergraduate or graduate students at a large U.S. university. Participants were paid $15 for their participation in the 1.5-hour experiment. Preliminary screening showed no significant demographic differences between participants in the two interface conditions.

**Materials**
*Serial Killer Task.* The task was based on a paradigm used in a number of previous studies of collaborative sensemaking [1, 2, 19, 21, 34, 35, 44]. In this task, each participant is provided with a set of documents pertaining to 3 cold murder cases, half of the documents pertaining to a current murder case, bus route information, and maps of the areas of the crimes. In total, there were seven murders, with about 40 potential suspects, hidden in about 20 documents divided equally between the two participants.

The task required participants to share their information in order to connect 10 clues spread across the cold cases and two extra clues in the unrelated current case. This combination of clues indicated that a serial killer was responsible for four of the cold cases and revealed the identity of that serial killer. In previous studies this task has proven to be quite difficult for participants such that the majority fail to identify the Serial Killer [e.g., 1, 2, 19, 21].

*Post-task report form.* After completing the task, participants were given individual paper report forms to complete. They were asked to provide the name of the serial killer, associated victims, and all clues that could incriminate the serial killer.

*Post-task survey.* An online post-task survey asked participants about their user experience and interface utility, collaboration experience, cognitive load (TLX), analytic ability, and demographic information. As described in more detail in the Measures section below, most questions were answered on 5 point Likert Scales.

**Equipment**
Two workstations (Intel Core i7 processor, 16 GB RAM) were connected to the Internet and ran SAVANT as a web application, deployed on the university server. Each was connected to two 25" monitors, the left showing the Document Space, and the right showing the Analysis Space. To simulate remote collaboration, the workstations were in separate cubicles to prevent eye contact and participants wore noise-cancelling headphones that prevented them from hearing their partner's speech or typing.

**Procedure**
Participants were seated apart at workstations such that they could not see each other or their partner's workstations. The experimenter explained that they would be role-playing detectives on a homicide team. After the participants signed the written consent form, they received training about the importance of *motive*, *opportunity*, and *lack of alibi* in solving crime cases. For experimenter's internal record keeping, participants with sensemaking translucent interface condition were assigned numbering in hundreds (100 onwards) and with control were assigned in tens (1 onwards). Next, they performed a 10-minute practice task in which they identified motive, opportunity and (lack of) alibi in a laptop theft crime case.

Next, participants received the instructions for collaborating on the crime task: to work together as a team, share information, and find the name of the serial killer. They were also given a demo of the SAVANT interface for their condition. Pairs were given 50 minutes to read through their documents, identify and share clues, brainstorm hypotheses, and identify the name of the serial killer. Upon completion of the task, they individually filled out the post-task report form and then the post-task survey.

## MEASURES

We have two main sources of data: participants' final reports, and post-task survey results.

### Task Performance

We used two measures for task performance, both based on the post task report form. Serial killer identification was a binary variable: 1 when correctly identified and 0 otherwise. Since this binary measure does not tell us how much progress a team had made in solving the case when the serial killer was not identified, we also used a clue recall score measured by the number of correct clues listed on the report form.

### Usefulness of Analysis Space

Participants responded to multiple questions in the post-task survey about the usefulness of the Analysis Space for spreading their attention across multiple cases, generating hypotheses, and collaborating on the task. These measures are based on those from other similar studies [7, 19, 51].

*Focused Attention Activity:* Five 5-point questions asked participants about the degree to which they interacted with the Analysis Space to pay attention to potential suspects, consider other alternative suspects, rule out suspects, track progress of suspects, and notice persons they did not pay enough attention to. For example, "I paid attention to number of potential suspects I considered in the Analysis Space". These five questions formed a reliable scale (Chronbach's α=.84) and were averaged to create a measure of Focused Attention.

*Hypothesis Activity*: Three 5-point questions asked participants about the degree to which the participants interacted with the Analysis Space to create hypotheses, confirm hypotheses and disconfirm hypotheses. These three questions formed a reliable scale (Chronbach's α=.71) and were averaged to create a measure of hypothesis Activity.

*Analysis Space Utility*: Five 5-point questions asked participants about the degree to which the Analysis Space helped them discuss cases with their partner, understand what their partner was thinking, track progress, and made them feel cognitively, and emotionally closer to their partner. These five questions formed a reliable scale (Chronbach's α=.88) and were averaged to create a measure of Analysis Space Utility.

### Team Experience

The post-task survey contained ten survey questions about the quality of the collaboration (e.g., "It was easy to discuss the cases with my partner," "My partner and I agreed about how to solve the case"). These ten questions formed a reliable scale (Cronbach's α=.84) and were averaged to create a team experience score, to answer H3. This measure is similar to [7, 19] who used a post-task questionnaire to assess quality of communication within the group.

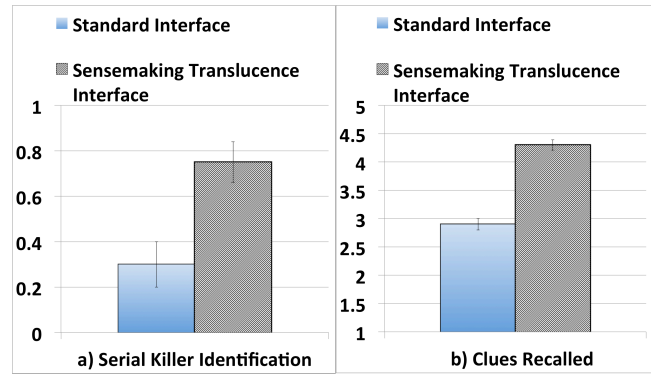

Figure 4. Serial killer identification and number of correct clues identified by interface condition.

### Cognitive Load

The post-task survey contained five questions based on the NASA TLX [26] that asked participants to rate how mentally demanding, temporally demanding, effortful, and frustrating the task was, as well as their subjective performance. After inverting the performance question, these five responses formed a reliable scale (Cronbach's α=.72). Participants' responses were averaged to create one measure of cognitive load.

## RESULTS

We present our findings in four sections. First, we discuss the effects of sharing sensemaking translucence on our two task performance measures. We then consider how it affected subjective ratings of SAVANT features, subjective ratings of how participants interacted with SAVANT, perceptions of team experience, and cognitive load.

### Task Performance

H1 proposed that pairs would perform better when sensemaking translucence was available than when it was not available. To test this hypothesis, we conducted mixed model ANOVAs, using clue recall and serial killer identification as our dependent measures. In these models, participant nested within pair was a random factor and interface condition (standard vs. sensemaking translucence) was a fixed factor.

*Clue recall.* There was a borderline significant effect of sensemaking translucence interface on the number of clues participants recalled in the written report (F[1, 38]=3.80, p=.06). As shown in Figure 4a, participants using the sensemaking translucence interface recalled more clues (M=4.3, SE=.47) than those using the standard interface (M=2.9, SE=.54).

*Serial Killer Identification.* Figure 4b shows participants' performance at identifying the name of the serial killer. Participants were significantly more likely to identify the name of the serial killer correctly when using the sensemaking translucence interface (M=.75, SE=.09) than when using the standard interface (M=.30, SE=.10; F[1, 38]=9.67, p=.004).

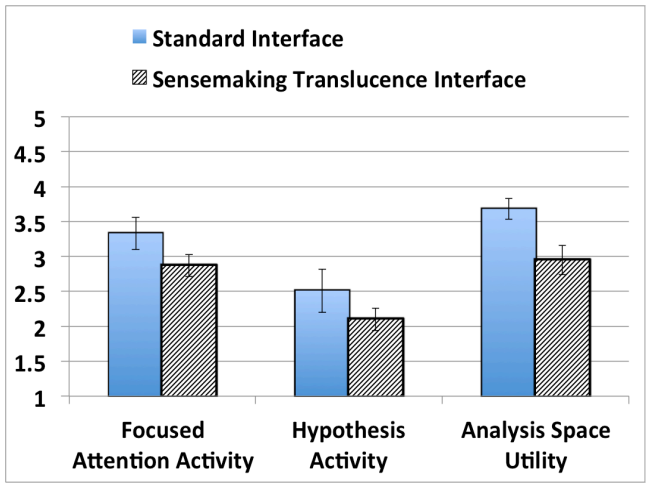

Figure 5. Perception of interface usefulness by interface condition

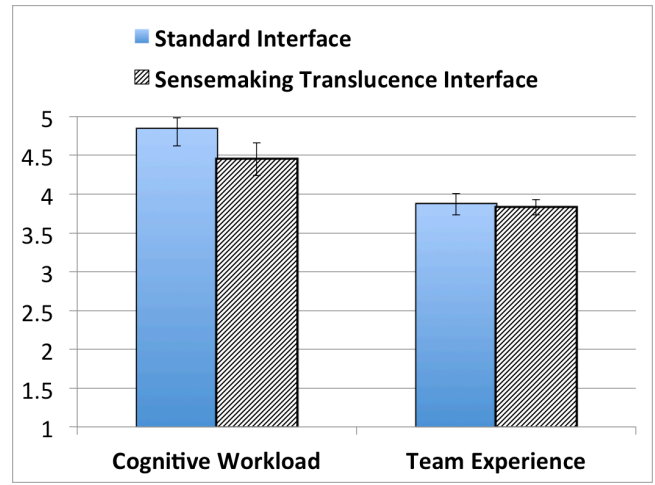

Figure 6. Self-reported workload and team experience by interface condition.

**Perception of Usefulness of SAVANT features**

According to H2a, sensemaking translucence would be perceived as more valuable. We analyzed participants' self-reported ratings of the user activity with SAVANT's features using mixed model ANOVAs with participants nested within pair as a random factor and interface condition (sensemaking translucence vs. standard) as a fixed factor.

Both focused attention activity and hypothesis activity (left two graphs in Figure 5) show a negative trend and did not support H2b. Participants in the sensemaking translucence condition reported using the Analysis Space to pay attention to potential suspects less (M=2.87, SE=.16) than did those without sensemaking translucence (M=3.33, SE=.23; F[1, 38]=2.56, p=.12). Participants in the sensemaking translucence condition also reported creating, confirming, and disconfirming hypothesis lesser (M=2.10, SE=.16) than those with no task-monitoring (M=2.51, SE=.31; F[1, 38]=1.39, p=.24). These results are opposite to H2b.

Further, the participants rated Analysis Space to be of lower utility when sensemaking translucence was available and did not support H2a (right graph in Figure 5). Participants in the non sensemaking translucence condition reported Analysis Space to be significantly better at helping them discuss cases and feel closer to their partner (M=3.68, SE=.15) than when sensemaking translucence was available (M=2.95, SE=.21; F[1, 38]=7.63, p =0.009).

**Team Experience**

H3 predicted that participants would rate the quality of their collaborations with their partners higher with sensemaking translucence compared to standard interface. To test this hypothesis, participants' team experience scores were analyzed in a mixed model ANOVA in which participant nested within pair was a random factor and condition (sensemaking translucence vs. standard) was a fixed factor. H3 was not supported (F[1, 38]=0.03, p=.84; see Figure 6).

**Cognitive Workload**

RQ1 asked whether cognitive workload would vary as a function of the presence or absence of sensemaking translucence. A mixed model ANOVA showed no significant difference between interface conditions (F[1, 38]=1.55, p=.21); participants with sensemaking translucence did not rate cognitive workload significantly lower (M=4.45, SE=0.21) than in the standard interface condition (M=4.84, SE=0.22). (See Figure 6.)

**Roles of Implicit and Explicit Sensemaking translucence**

Participants' open-ended responses on the post-survey provided details about how the different features in Analysis Space were appropriated to collect clues, and solve the cases.

*Hypothesis Window.* First, several participants mentioned the interplay between the implicitly shared stickies and the Hypothesis Window. They referred to the bidirectional knowledge transfer between these two channels and showed how the two complemented each other:

*"After deciding on an MO [Modus Operandus] for the serial killer in the hypothesis space, we moved that out of the window and onto a stickie."* – P111, female

*"We used the analysis space to connect sticky notes and then form hypotheses based on the notes"…"We more sketched out ideas in the hypothesis space I think, after we had agreed on them"* – P122, female

*"The hypothesis window was very useful for synthesizing all the evidence we found in an easy-to-read window so that we could keep track of our findings. The stickies provided all the supporting evidence, but the hypothesis window summarized it for us."* – P106, male

Several participants also mentioned the partial use of the Hypothesis Window. Most participants created hypothesis towards the second-half of investigation, unless they were confident. Further, most participants also reported creating or supporting previously created hypothesis and shied away from adding evidences that disconfirmed hypothesis that had been decided upon previously with their partner:

*"I created hypothesis for my cases with common MO. I confirmed the hypothesis made by me and the other person on our Current case. Didn't disconfirm the other person's hypothesis due to lack of detailed information about his/her cold cases." – P103, female*

*"Used the hypothesis window by including supporting evidence only. I didn't write any of the hypotheses only my partner did. Didn't include any evidence to reject the hypothesis… " – P128, male*

As is evident, much of the value of Hypothesis Window came from hypothesis reporting, evidence gathering, hypothesis confirmation, and combination with stickies in analysis space. Other participants shared their strategies of how they optimized use of the analysis space, and pointed to two distinct advantages of Hypothesis Window. First, it enables better organization of ideas than the free form stickies. Second, it was used not to report the emergent hypothesis but conclusions, instead.

*"I found that I was able to comprehend all the evidence best by looking at all of our stickies, setting up our hypotheses, and talking about what we thought. However, I think in order for stickies to be useful they have to be organized nicely. Sometimes when my partner's stickies were disorganized I had a hard time following her cases." – P106, male*

*"Being able to chat was very important for talking about current ideas. Stickies helped to identify important things, but it was easy to overload on stickies, especially with 6 cases in one space. The hypothesis window was good for when we already felt we had conclusions, but did not necessarily come in handy during the thought process. The visualization was good for seeing where we needed to do more work but not really good for focusing on people." – P111, female*

On the contrary, pairs without the Hypothesis Window also used both implicit and explicit channels, in agreement with previous research [19]:

*"By using notes about each case and comparing them to one another." – P14, female*

*"I wrote any details that could possibly mean any type of link (ie: someone worked with money, someone had a secret lover, someone worked in the same field as another victim)." - P18, female*

*Suspect Visualization.* Unlike the Hypothesis Window, participants reported using the Suspect Visualization for not just confirming, but also disconfirming their hypotheses. Several participants reported using the visualization to ensure that they did not focus attention on any particular case/suspect:

*"Paid attention to the avatars mainly to make sure that I didn't concentrate on one case only" – P103, female*

*"I only paid attention to the avatars if we were talking too much about a certain person. I didn't think until the end to consider it for potential suspects." – P115, male*

Further, the visualization helped indicate lack of sufficient information about potential suspects, so that they could not be discounted. So, instead of removing potential suspects, visualization helped users to rule in potential suspects:

*"I didn't rule out suspects due to the visualization but I did use the visualization to see if I needed more information on a suspect." – P111, female*

Avatars in the visualization also served to imply whether the pairs shared common ground about the potential correct answer by showing how much the collaborators were referring to any particular suspect together:

*"I used the avatars to let me know that there was a certainty that my partner and I were on the same page about suspecting someone for the crime" – P109, male*

For some others, the visualization either served as tool for confirming their "hunch" or had no effect on their sensemaking process:

*"I hardly glanced at it. There were too many people for me to start accounting for them all. I only started looking into a suspect when I began noticing where they fit into the "story"." – P106, male*

*"Mentioning the name didn't really change my investigation of them or others didn't affect my work didn't rule anyone out" – P104, female*

A few participants reported visualization blindness because visualizing suspects from previously reviewed cases could be irrelevant to newer threads of investigation.

*"There were too many avatars for me to keep track of everyone, and we were on a time crunch. I don't think it is worth investigating someone more just because we haven't talked about them much. If they have a solid alibi, or are involved in a case that has nothing in common with our case, they are not as important as the person who is mentioned across several files." – P106, male*

*"I didn't really use the Avatars and visualization at the bottom since I felt that it didn't really help and that it was there just as a distraction" – P110, female*

On the other hand, other participants found value in the visualization by optimizing strategies, instead of completely ignoring it. Some participants did not view the Avatars, unless they visualized sensemaking translucence on the

"local" current thread of investigation, instead of "global" sensemaking translucence:

*"I didn't use the avatars if they did not have to do with the case I was working on at the time. When I made new notes about small details, it added every single name I mentioned although it didn't have to do with another case I was working on." – P109, male*

Others found value in the visualization as sources for new investigation-threads and hypothesis based on the names, or how relevant these names seemed to be potential suspects:

*"The avatars in the beginning were adding up very quickly. Not until two-thirds of the way into the time did certain avatars seem like true suspects to consider. And only then did my partner and I start developing hypotheses." – P128, male*

In addition to the quantitative results, these comments show how the Hypothesis Window was viewed as less useful than we expected. Even though participants liked its structured organization, they used it to primarily report and confirm conclusions instead of reporting emerging hypothesis and then working to confirm and disconfirm them. The Suspect Visualization was viewed as useful to identify potential suspects and validate common-ground. While both sensemaking translucence features connect with communication channels in the Analysis Space, the visualization improved collaborative analysis by creating common ground and initiating new threads of investigation.

**DISCUSSION**

We studied the impact of sensemaking translucence in a distributed synchronous collaborative sensemaking task. Consistent with H1, we found that pairs of analysts using an interface that provided sensemaking translucence identified significantly more clues relevant to solving the case and identified the culprit a significantly greater proportion of the time than pairs using the standard interface with no sensemaking translucence features. However, inconsistent with our other hypotheses, pairs using the sensemaking translucence interface rated it as less helpful than pairs using the standard interface in terms of providing support for analysis, aiding hypothesis generation, and helping them pay attention to multiple suspects.

There could be multiple reasons for this mismatch between the objective value of the tool for task performance and users' subjective perceptions of its value. First, participants may have been uncomfortable with the amount of explicit sharing of information and insights required in the sensemaking translucence interface even though this greater explicitness helped them solve the case. The sensemaking translucence interface required explicit actions by the users in terms of when and what to write into the Hypothesis Window. This tension between explicit sharing for sensemaking translucence vs. ease of use could explain lower subjective ratings of the sensemaking translucence interface. Previous work shows that implicit sharing leads to significantly better user experience compared to when explicit actions are required by the collaborators [19].

Second, participants had to spread their attention across far more features and information made available by these features in sensemaking translucence over the standard feature, even though this information helped them identify potential suspects. As P110 and P106 point out, participants using the sensemaking translucence interface had to distribute their attention between stickies, chat, an organized hypothesis window, and the suspect visualization whereas participants using the standard interface had only to attend to the stickies and chat. While the former did not lead to a significantly higher level of self-reported cognitive load, it may be that at some level participants reacted negatively to having to manage so many different features at once, resulting in lower reported perceived utility.

Third, despite the obvious gains in task performance, tools like our hypothesis window that enforce rigid structure are often perceived to be of low utility in collaborative sensemaking tasks. For example, Convertino et al [10]. found that almost half of their participants wanted more control of how information was displayed in a matrix similar to SAVANT's Hypothesis Window. As P122 and P106 point out, even though the structured hypothesis window provided a strict organization to conclusions, the unstructured nature of the stickies allowed them to sketch out ideas prior to inserting their conclusions in the hypothesis window. However, previous research indicates that despite the lack of *perceived* utility, rigid structure improves task performance in decision-making groups [37].

Fourth, we believe that factors such as hindsight bias [48], may lead users to misattribute their success to their own cognitive abilities rather than the information made available by the features of our sensemaking translucence interface. While users were more likely to identify the serial killer with sensemaking translucence, they may have attributed success to themselves as opposed to the interface.

In addition to the incongruence between our results for task performance and perceived usability, we also found that participants appropriated the sensemaking features differently because the two features were designed to support different parts of the foraging and the sensemaking process. While the Hypothesis Window was designed to encourage sensemaking through explicit confirmation/disconfirmation of partners' insights, the Suspect Visualization was designed to discourage focusing attention on any particular suspect early in the sensemaking process. In the next paragraphs, we discuss how our participants appropriated each of these features for sensemaking.

The Hypothesis Window was appropriated as a summarizing tool. One of the reasons could be that the participants may have to balance between the immediacy of hypothesis/evidence reporting with the perceived benefit of leaving their current sensemaking loop to do so. Some participants (P103 and P128) mentioned that they filled in the Hypothesis Window only towards the end of the process, rather than in an ongoing fashion as intended. This suggests a need to pay greater attention to the temporal nature of people's use of sensemaking translucence features. This connects well with Reddy's discussion of rhythms of work in information seeking [44], which suggests that individuals' actions are dependent not just on immediacy but also on when it would be most beneficial to perform them in their work.

The Suspect Visualization was appropriated for foraging to rule in potential suspects. One of the reasons could be that the visualization affords an overall view of sensemaking across the 6 cases, instead of specifically supporting sensemaking of a single case. Users pointed to sensemaking as an act of pursuing multiple threads of investigation: global and local. As suggested by some participants (P106 and P109), visualizing the "average" global attention-spread prevents depicting the attention-spread at a local case level, possibly leading them to view the Suspect Visualization as "distracting" when a pattern of a serial nature is not yet evident.

**LIMITATIONS**

This study is about a time-limited synchronous crime-solving collaborative tasks. Further work is needed to clarify the value of sensemaking translucence for other types of tasks. This study also focused on two possible designs in a large design space for collaborative analysis tools, and it studied the impact of using the two features together instead of assessing the cost vs. benefit of each feature individually. More research will be needed to determine the best possible way to implement these features in collaborative sensemaking tools. Further, while this study focused on a study between a pair of collaborators, future research is needed to understand the impact of team size, and scaling across multiple collaborators [23, 24] on designing sensemaking translucence features.

In addition, while lab-settings enable designers to vary design choices in controlled settings to understand the effects of each choice across multiple measures, field research is needed to understand the impact of sensemaking translucence in real life crime solving teams analyzing datasets of varying sizes. One particular challenge might be the scalability of sensemaking translucence. With increasing time and complexity of datasets, or with increasing team size, explicitly sharing at a suspect and hypothesis level could lead to information overload. At the same time, however, it might provide data provenance that is often lost in sensemaking. Further, domains like crime-solving/medical-sensemaking etc. are constrained in the extent to which certain types of information can be shared due to organizational privacy laws. We see this as an open design space, where designers could consider single or multiple means of visualizing the sensemaking translucence.

**DESIGN IMPLICATIONS**

The findings of our study have implications for designing synchronous distributed collaborative sensemaking tools. Since the presence of sensemaking translucence had a significant positive effect on task performance, we propose that tools should support sensemaking translucence to promote timely information sharing and reduced confirmation bias. However, the tension between non-premature hypothesis sharing, and timely information sharing should be reduced. We see this as a spectrum where designers could design for fully automated hypothesis generation or require limited input from users to verify generated hypothesis, or any solution in between. Designers could push natural language processing further to identify not just the named-entities (suspects) but the explicitly shared hypothesis too. Identified hypotheses may be implicitly shared using the hypotheses window, reducing the redundancy and improving sensemaking translucence. Improved sensemaking translucence without explicit sharing would reduce workload and will make user attention available for the process of sensemaking itself.

Providing interfaces that free up user attention could also help improve subjective user experience while maintaining high task performance. Connecting sensemaking translucence more closely to the artifacts of sensemaking itself could also reduce this tension between user-experience and task performance. The Hypothesis Window, and Suspect Visualization could be better integrated with the stickies to further accrue the advantages of implicit sharing. For example, evidence that disconfirms any hypothesis could be highlighted for closer inspection. Eventually, collaborators could leverage each other's insights as recommended by the system while they manage their limited attention under strict time pressure.

Future designs would also benefit from greater attention to how sensemaking features might support the different phases of sensemaking work itself. One design goal could be to remind users to interact with sensemaking translucence features more during the process of sensemaking. Future tools could match natural language processing on the chat transcripts with machine learning on user activity for behavior recognition. Recognizing activities correlated with task success vs. failure could help customize the tool usage, for instance by recommending to users that they need to pursue disconfirmation of existing hypothesis or that they should distribute their attention across suspects. Detecting deviations from known successful behaviors and persuading investigators to reduce biases could potentially reduce the number of unsolved cases [43].

Finally, more closely integrated sensemaking translucence features could help reduce workload. In our design, integration was unidirectional such that the Hypothesis

Window drove the visualization but not vice versa. It is also possible for the visualization to trigger new hypotheses about potential criminals. Implicit hypotheses (including location, timeline, alibi etc.) could be generated based on notes gathered for each suspect. This might aid sensemaking translucence by offering an alternative bottom-up view.

**CONCLUSION**

In this paper, we presented findings from an experiment in which pairs of participants played the role of crime analysts collaborating to identify a serial killer in a distributed synchronous setting. We found that implicit sharing aided by sensemaking translucence improved both clue-detection and success at identifying the hidden serial killer, without increasing cognitive workload. Despite significantly higher task performance, users reported lower utility of the interface and no increased sense of success in the collaboration when using the sensemaking translucence interface. While, sensemaking translucence may enable analysts to perform better in such situations, further research is needed to ensure that analysts themselves perceive the interface as useful.

**REFERENCES**

1. Aruna D. Balakrishnan, Susan R. Fussell, and Sara Kiesler. 2008. Do visualizations improve synchronous remote collaboration? In *Proceedings of the* SIGCHI *Conference on Human Factors in Computing Systems* (CHI '08). ACM, New York, NY, USA, 1227-1236. DOI=10.1145/1357054.1357246 http://doi.acm.org/10.1145/1357054.1357246

2. Aruna D. Balakrishnan, Susan R. Fussell, Sara Kiesler, and Aniket Kittur. 2010. Pitfalls of information access with visualizations in remote collaborative analysis. In *Proceedings of the 2010 ACM conference on Computer supported cooperative work* (CSCW '10). ACM, New York, NY, USA, 411-420. DOI=10.1145/1718918.1718988 http://doi.acm.org/10.1145/1718918.1718988

3. Jakob E. Bardram, Thomas R. Hansen, and Mads Soegaard. 2006. AwareMedia: a shared interactive display supporting social, temporal, and spatial awareness in surgery. In *Proceedings of the 2006 20th anniversary conference on Computer supported cooperative work* (CSCW '06). ACM, New York, NY, USA, 109-118. DOI=10.1145/1180875.1180892 http://doi.acm.org/10.1145/1180875.1180892

4. A. J. Bernheim, David Bargeron, Jonathan Grudin, and Anoop Gupta. 2002. Notification for shared annotation of digital documents. In *Proceedings of the SIGCHI Conference on Human Factors in Computing Systems* (CHI '02). ACM, New York, NY, USA, 89-96. DOI=10.1145/503376.503393 http://doi.acm.org/10.1145/503376.503393

5. Eric A. Bier, Stuart K. Card, and John W. Bodnar, Principles and Tools for Collaborative Entity-Based Intelligence Analysis, *IEEE Transactions on visualization and Computer Graphics*, v.16 n.2, p.178-191, March 2010. http://doi.acm.org/10.1109/TVCG.2009.104

6. George Chin, Jr., Olga A. Kuchar, and Katherine E. Wolf. 2009. Exploring the analytical processes of intelligence analysts. In *Proceedings of the SIGCHI Conference on Human Factors in Computing Systems* (CHI '09). ACM, New York, NY, USA, 11-20. DOI=10.1145/1518701.1518704 http://doi.acm.org/10.1145/1518701.1518704

7. Mei C. Chuah, and Steven F. Roth. 2003. Visualizing Common Ground. In *Proceedings of the Seventh International Conference on Information visualization* (IV '03). IEEE Computer Society, Washington, DC, USA, 365-.

8. Haeyong Chung, Seungwon Yang. Naveed Massjouni, Christopher Andrews, Rahul Kanna, and Chris North. 2010. Vizcept: Supporting synchronous collaboration for constructing visualizations in intelligence analysis. In *Proceedings of VAST '10,* 107--114.

9. David Constant, Sara Kiesler, and Lee Sproull. 1994. What's mine is ours, or is it? A study of attitudes about information sharing. *Info. Systems Research*, 5, 400--421.

10. Gregorio Convertino, Dorrit Billman, Peter Pirolli, J. P. Massar, and Jeff Shrager. 2008. The CACHE Study: Group Effects in Computer-supported Collaborative Analysis. *Comput. Supported Coop. Work* 17, 4 (August 2008), 353-393. DOI=10.1007/s10606-008-9080-9 http://dx.doi.org/10.1007/s10606-008-9080-9

11. Gregorio Convertino, Helena M. Mentis, Mary Beth Rosson, Aleksandra Slavkovic, and John M. Carroll. 2009. Supporting content and process common ground in computer-supported teamwork. In *Proceedings of the SIGCHI Conference on Human Factors in Computing Systems* (CHI '09). ACM, New York, NY, USA, 2339-2348. DOI=10.1145/1518701.1519059 http://doi.acm.org/10.1145/1518701.1519059

12. Gregorio Convertino, Helena M. Mentis, Aleksandra Slavkovic, Mary Beth Rosson, and John M. Carroll. 2011. Supporting common ground and awareness in emergency management planning: A design research project. In *ACM Trans. Comput.-Hum. Interact.* 18, 4, Article 22 (December 2011), 34 pages. DOI=10.1145/2063231.2063236 http://doi.acm.org/10.1145/2063231.2063236

13. Steven A. Egger. 1998. *The killers among us: An examination of serial murder and its investigation*. Upper Saddle River, NJ: Prentice Hall.

14. Kristie Fisher, Scott Counts, and Aniket Kittur. 2012. Distributed sensemaking: improving sensemaking by leveraging the efforts of previous users. In *Proceedings of the SIGCHI Conference on Human Factors in*

*Computing Systems* (CHI '12). ACM, New York, NY, USA, 247-256. DOI=10.1145/2207676.2207711 http://doi.acm.org/10.1145/2207676.2207711

15. Jon Froehlich, and Paul Dourish. 2004. Unifying Artifacts and Activities in a Visual Tool for Distributed Software Development Teams. In *Proceedings of the 26th International Conference on Software Engineering* (ICSE '04). IEEE Computer Society, Washington, DC, USA, 387-396.

16. Craig H. Ganoe, Jacob P. Somervell, Dennis C. Neale, Philip L. Isenhour, John M. Carroll, Mary Beth Rosson, and D. Scott McCrickard. 2003. Classroom BRIDGE: using collaborative public and desktop timelines to support activity awareness. In *Proceedings of the 16th annual ACM symposium on User interface software and technology* (UIST '03). ACM, New York, NY, USA, 21-30. DOI=10.1145/964696.964699 http://doi.acm.org/10.1145/964696.964699

17. Darren Gergle, Robert E. Kraut, and Susan R. Fussell. 2004. Language efficiency and visual technology minimizing collaborative effort with visual information. *Journal of Language and Social Psychology* 23: 491-517.

18. Steven Gottlieb. 1994. *Crime analysis: From first report to final arrest*. Montclair, CA: Alpha Publishing.

19. Nitesh Goyal, and Susan R. Fussell. 2015. Designing for Collaborative Sensemaking: Leveraging Human Cognition For Complex Tasks. In *Proceedings of the IFIP TC.13 INTERACT 2015 conference*. Springer

20. Nitesh Goyal, Gilly Leshed, Dan Cosley, and Susan R. Fussell. 2014. Effects of implicit sharing in collaborative analysis. In *Proceedings of the SIGCHI Conference on Human Factors in Computing Systems* (CHI '14). ACM, New York, NY, USA, 129-138. DOI=10.1145/2556288.2557229 http://doi.acm.org/10.1145/2556288.2557229

21. Nitesh Goyal, Gilly Leshed, and Susan R. Fussell. 2013. Effects of visualization and note-taking on sensemaking and analysis. In *Proceedings of the SIGCHI Conference on Human Factors in Computing Systems* (CHI '13). ACM, New York, NY, USA, 2721-2724. DOI=10.1145/2470654.2481376 http://doi.acm.org/10.1145/2470654.2481376

22. Nitesh Goyal, Gilly Leshed, and Susan R. Fussell. 2013. Leveraging partner's insights for distributed collaborative sensemaking. In *Proceedings of the 2013 conference on Computer supported cooperative work companion* (CSCW '13). ACM, New York, NY, USA, 15-18. DOI=10.1145/2441955.2441960 http://doi.acm.org/10.1145/2441955.2441960

23. Nitesh Goyal. 2015. Designing for Collaborative Sensemaking: Leveraging Human Cognition For Complex Tasks. In *Extended Abstracts of the 15th IFIP TC.13 International Conference on Human-Computer Interaction INTERACT 2015*.

24. Nitesh Goyal. 2015. Designing for Collaborative Sensemaking: Using Expert & Non-Expert Crowds. In *Extended Abstracts of Third AAAI Conference on Human Computation and Crowdsourcing* HCOMP, 2015

25. Carl Gutwin, and Saul Greenberg. 1998. Design for individuals, design for groups: tradeoffs between power and workspace awareness. In *Proceedings of the 1998 ACM conference on Computer supported cooperative work* (CSCW '98). ACM, New York, NY, USA, 207-216. DOI=10.1145/289444.289495 http://doi.acm.org/10.1145/289444.289495

26. Sandra G. Hart, and Lowell E. Staveland. 1988. Development of a multi-dimensional workload rating scale. In *P. A. Hancock and N. Mesh Kati (Eds.), Human mental workload*, 139--183. Amsterdam: Elsevier.

27. Stephen C. Hayne, Lucy J. Troup, and Sara A. Mccomb. 2011. "Where's Farah?": Knowledge silos and information fusion by distributed collaborating teams. *Information Systems Frontiers* 13, 1 (March 2011), 89-100. DOI=10.1007/s10796-010-9274-9 http://dx.doi.org/10.1007/s10796-010-9274-9

28. Jeffrey Heer, and Maneesh Agrawala. 2008. Design considerations for collaborative visual analytics. *Information visualization* 7, 1 (March 2008), 49-62. DOI=10.1145/1391107.1391112 http://dx.doi.org/10.1145/1391107.1391112

29. Jeffrey Heer, Fernanda B. Viégas, and Martin Wattenberg. 2009. Voyagers and voyeurs: Supporting asynchronous collaborative visualization. *Commun. ACM* 52, 1 (January 2009), 87-97. DOI=10.1145/1435417.1435439 http://doi.acm.org/10.1145/1435417.1435439

30. Richards J. Heuer Jr., 1999. *The psychology of intelligence*. Washington DC: Center for the Study of Intelligence, Government Printing Office.

31. Petra Isenberg, Danyel Fisher, Sharoda A. Paul, Meredith Ringel Morris, Kori Inkpen, and Mary Czerwinski. 2012. Co-Located Collaborative Visual Analytics around a Tabletop Display. *IEEE Transactions on visualization and Computer Graphics* 18, 5 (May 2012), 689-702. DOI=10.1109/TVCG.2011.287 http://dx.doi.org/10.1109/TVCG.2011.287

32. Jeroen Janssen, Gijsbert Erkens, Gellof Kanselaar, and Jos Jaspers. 2007. Visualization of participation: Does it contribute to successful computer-supported collaborative learning? *Comput. Educ.* 49, 4 (December 2007), 1037-1065. DOI=10.1016/j.compedu.2006.01.004 http://dx.doi.org/10.1016/j.compedu.2006.01.004

33. Sirkka Jarvenpaa, and D. S. Staples 2000. The use of collaborative electronic media for information sharing: An exploratory study of determinants. *The Journal of Strategic Information Systems*, 9(2), 129--154.

34. Ruogu Kang, Aimee Kane, and Sara Kiesler. 2014. Teammate inaccuracy blindness: when information sharing tools hinder collaborative analysis. In *Proceedings of the 17th ACM conference on Computer supported cooperative work & social computing* (CSCW '14). ACM, New York, NY, USA, 797-806. DOI=10.1145/2531602.2531681 http://doi.acm.org/10.1145/2531602.2531681

35. Ruogu Kang, and Sara Kiesler. 2012. Do collaborators' annotations help or hurt asynchronous analysis. In *Proceedings of the ACM 2012 conference on Computer Supported Cooperative Work Companion* (CSCW '12). ACM, New York, NY, USA, 123-126. DOI=10.1145/2141512.2141558 http://doi.acm.org/10.1145/2141512.2141558

36. JinKyu Lee, and H. Raghav Rao. Exploring the causes and effects of inter-agency information sharing systems adoption in the anti/counter-terrorism and disaster management domains. In *Proceedings of the 8th annual international conference on Digital government research: bridging disciplines & domains*, May 20-23, 2007, Philadelphia, Pennsylvania

37. Helena M. Mentis, Paula M. Bach, Blaine Hoffman, Mary Beth Rosson, and John M. Carroll. 2009. Development of decision rationale in complex group decision making. In *Proceedings of the SIGCHI Conference on Human Factors in Computing Systems* (CHI '09). ACM, New York, NY, USA, 1341-1350. DOI=10.1145/1518701.1518904 http://doi.acm.org/10.1145/1518701.1518904

38. National Commission on Terrorist Attacks upon the United States, 2004. *The 9/11 Commission Report: Final report of the national commission on terrorist attacks upon the United States*, Norton, NY

39. Syavash Nobarany, Mona Haraty, and Brian Fisher. Facilitating the reuse process in distributed collaboration: a distributed cognition approach. In *Proceedings of the ACM 2012 conference on Computer Supported Cooperative Work*, (CSCW '12). February 11-15, 2012, Seattle, Washington, USA. DOI=10.1145/2145204.2145388 http://doi.acm.org/10.1145/2145204.2145388

40. Sharoda A. Paul, and Madhu C. Reddy. 2010. Understanding together: sensemaking in collaborative information seeking. In *Proceedings of the 2010 ACM conference on Computer supported cooperative work* (CSCW '10). ACM, New York, NY, USA, 321-330. DOI=10.1145/1718918.1718976 http://doi.acm.org/10.1145/1718918.1718976

41. Nicholas J. Pioch, and John O. Everett. 2006. POLESTAR: collaborative knowledge management and sensemaking tools for intelligence analysts. In *Proceedings of the 15th ACM international conference on Information and knowledge management* (CIKM '06). ACM, New York, NY, USA, 513-521. DOI=10.1145/1183614.1183688 http://doi.acm.org/10.1145/1183614.1183688

42. Peter Pirolli, and Stuart Card. 2005. The sensemaking process and leverage points for analyst technology as identified through cognitive task analysis. *Proc. ICA* Vol. 5, 2—4

43. *Police Chief magazine*, October 2009 http://www.policechiefmagazine.org/magazine/index.cfm?fuseaction=display_arch&article_id=1922&issue_id=102009

44. Madhu Reddy and Paul Dourish. "A finger on the pulse: temporal rhythms and information seeking in medical work." Proceedings of the 2002 *ACM conference on Computer supported cooperative work.* ACM, 2002.

45. Peter Scupelli, Susan R. Fussell, Sara Kiesler, Pablo Quinones, and Gail Kusbit. 2007. Juggling Work Among Multiple Projects and Partner. In *Proceedings of the 40th Annual Hawaii International Conference on System Sciences* (HICSS '07). IEEE Computer Society, Washington, DC, USA, 77-. DOI=10.1109/HICSS.2007.310 http://dx.doi.org/10.1109/HICSS.2007.310

46. John Stasko, Carsten Gorg, and Zhicheng Liu. 2008. Jigsaw: supporting investigative analysis through interactive visualization. *Information visualization* 7, 2 (April 2008), 118-132. DOI=10.1145/1466620.1466622 http://dx.doi.org/10.1145/1466620.1466622

47. Anthony Tang, Melanie Tory, Barry Po, Petra Neumann, and Sheelagh Carpendale. 2006. Collaborative coupling over tabletop displays. In *Proceedings of the SIGCHI Conference on Human Factors in Computing Systems* (CHI '06), Rebecca Grinter, Thomas Rodden, Paul Aoki, Ed Cutrell, Robin Jeffries, and Gary Olson (Eds.). ACM, New York, NY, USA, 1181-1190. DOI=10.1145/1124772.1124950 http://doi.acm.org/10.1145/1124772.1124950

48. Amos Tversky and Daniel Kahneman. "Judgment under uncertainty: Heuristics and biases." *science* 185.4157 (1974): 1124-1131.

49. Chris Weaver. 2007. Is Coordination a Means to Collaboration? In *Proceedings of the Fifth International Conference on Coordinated and Multiple Views in Exploratory visualization* (CMV '07). IEEE Computer Society, Washington, DC, USA, 80-84. DOI=10.1109/CMV.2007.15 http://dx.doi.org/10.1109/CMV.2007.15

50. Wesley Willett, Jeffrey Heer, Joseph Hellerstein, and Maneesh Agrawala. 2011. CommentSpace: structured support for collaborative visual analysis. In *Proceedings of the SIGCHI Conference on Human Factors in Computing Systems* (CHI '11). ACM, New York, NY, USA, 3131-3140. DOI=10.1145/1978942.1979407 http://doi.acm.org/10.1145/1978942.1979407

51. William Wright, David Schroh, Pascale Proulx, Alex Skaburskis, and Brian Cort. 2006. The Sandbox for analysis: concepts and methods. In *Proceedings of the SIGCHI Conference on Human Factors in Computing Systems* (CHI '06), Rebecca Grinter, Thomas Rodden, Paul Aoki, Ed Cutrell, Robin Jeffries, and Gary Olson (Eds.). ACM, New York, NY, USA, 801-810. DOI=10.1145/1124772.1124890 http://doi.acm.org/10.1145/1124772.1124890